\documentclass{aa} 

\usepackage{graphicx}

\usepackage{txfonts}

\usepackage{natbib,twoopt}
\usepackage{array}
\usepackage{enumitem} 
\usepackage{threeparttable}
\usepackage[breaklinks=true, colorlinks=true, linkcolor=blue, citecolor=blue, urlcolor=blue]{hyperref}

\begin{document}

   \title{Impact of convective overshooting on the single-degenerate model of Type Ia supernovae}

  \titlerunning{Impact of convective overshooting on the SD model of SNe Ia} 
   \author{Xinghuan Wang 
          \inst{1,2},
          Xiangcun Meng
          \inst{1}
          }

          \institute{International Centre of Supernovae (ICESUN), Yunnan Key Laboratory of Supernova Research, Yunnan Observatories, Chinese Academy of Sciences (CAS), Kunming 650216, China\\
          \email{wangxinghuan@ynao.ac.cn, xiangcunmeng@ynao.ac.cn}
          \and University of Chinese Academy of Sciences, Beijing 100049, PR China}

  \abstract
   {
    The single-degenerate (SD) model is one of the principal models for the progenitors of Type Ia supernovae (SNe Ia).  
    However, it faces some challenges, the primary being its inability to account for the observed SN Ia birth rate.   
    Many studies have attempted to address this issue by expanding the parameter space, defined by the initial donor star mass and orbital period, that can lead to SNe Ia, as well as by improving binary population synthesis.
    While these efforts have led to significant progress, many uncertainties in stellar physics persist, which influences the outcomes of such studies.
    Convective overshooting, which can significantly affect the internal structure of a star and subsequently its evolution within a binary system, is one of the most significant sources of uncertainty in stellar physics.
   }
   {
    We investigate the effect of convective overshooting on the parameter space and birth rate of SNe Ia within the SD model.
   }
   { We employed the common-envelope wind (CEW) model, a new version of the SD model, as our progenitor model. 
    Using MESA, we obtained the parameter space that leads to SNe Ia for three different convective overshooting parameters and calculated the corresponding SN Ia birth rate.
    }
   {
    Convective overshooting expands the upper boundaries (corresponding to a larger initial donor mass) and right boundaries (corresponding to a longer initial orbital period) of the parameter space for systems with massive white dwarfs (WDs; $\ge$ 0.75 $M_\mathrm{\odot}$).     
    However, the minimum WD mass and the parameter space for low-mass WDs - and, consequently, the calculated SN Ia birth rate - vary non-monotonically with convective overshooting parameters.
    The CEW model may explain the SNe Ia that interact with the circumstellar medium (CSM), i.e., SNe Ia-CSM. 
    We find that the parameter space for SNe Ia-CSM increases with convective overshooting parameters, as does their birth rate.
   }
  {}


   \maketitle

\section{Introduction}\label{sec:1}

Type Ia supernovae (SNe Ia) are excellent cosmological distance indicators that enable the measurement of cosmological parameters and thus can be used to determine the accelerating expansion of the Universe and whether dark energy exists (e.g., \citealt{1998AJ....116.1009R}; \citealt{1999ApJ...517..565P}; \citealt{2011NatCo...2..350H}; \citealt{2015IJMPD..2430029M}). 
SNe Ia are crucial to understanding the evolution of galactic chemical compositions because they are the primary source of iron production in the Universe (e.g., \citealt{1983A&A...118..217G}; \citealt{1986A&A...154..279M}). 
It is widely accepted that a SN Ia originates from the thermonuclear explosion of a carbon-oxygen white dwarf (CO WD) in binary systems. Both the Chandrasekhar mass and the sub-Chandrasekhar mass models are theoretically capable of producing explosions (e.g., \citealt{1960ApJ...132..565H}; \citealt{2017MNRAS.472.2787N}; \citealt{2025A&ARv..33....1R}). 
Based on the different properties of the companion star, progenitor models for SNe Ia are primarily divided into two categories: the double-degenerate (DD) model and the single-degenerate (SD) model (e.g., \citealt{2008NewAR..52..381P}; \citealt{2012NewAR..56..122W}; \citealt{2014ARA&A..52..107M}; \citealt{2014NewAR..62...15R}; \citealt{2019NewAR..8701535S}; \citealt{2023RAA....23h2001L}). 
In the DD model, two WDs can produce a SN Ia when they merge (e.g., \citealt{1984ApJS...54..335I}; \citealt{1984ApJ...277..355W}; \citealt{1998MNRAS.296.1019H}; \citealt{2001A&A...365..491N}; \citealt{2012A&A...546A..70T}). 
In the SD model, a CO WD accretes hydrogen-rich or helium-rich material from a non-degenerate companion, which could be a main-sequence (MS) star, a red giant (RG) star, or a helium (He) star (e.g., \citealt{1973ApJ...186.1007W};  \citealt{1984ApJ...286..644N}; \citealt{1997A&A...322L...9L}; \citealt{2000A&A...362.1046L}; \citealt{2004MNRAS.350.1301H}; \citealt{2007ApJ...658L..51C}; \citealt{2009MNRAS.395..847W}; \citealt{2011ApJ...735L..31C}; \citealt{2024ApJ...974...13L}).
Both models have some supporting evidence, but both also face significant issues in terms of observations and theory. 

In our study we focus on the SD model. The detection of circumstellar material around SNe Ia provides strong evidence for the SD scenario (e.g., \citealt{2013MNRAS.436..222M}; \citealt{2019MNRAS.487.2372V}; \citealt{2023ApJ...944..204U}), as does the early optical and ultraviolet emission resulting from ejecta--companion interactions observed in some SNe Ia (e.g., \citealt{2011MNRAS.416.2607G}; \citealt{2015Natur.521..328C}; \citealt{2015MNRAS.454.1192L}; \citealt{2016ApJ...826...96P}).

However, the SD model faces some challenges, the primary being that the CO WD accretion rate must lie within a narrow range for a stable burning of accreted material, which makes it difficult to account for the observed SN Ia birth rate in the Galaxy (e.g., \citealt{2012NewAR..56..122W}; \citealt{2014ARA&A..52..107M}; \citealt{2014NewAR..62...15R}; \citealt{2018RAA....18...49W}; \citealt{2019NewAR..8701535S}; \citealt{2023RAA....23h2001L}). 
Many studies have attempted to address this issue by expanding the parameter space, defined by the initial donor star mass and orbital period, that can lead to SNe Ia (e.g., \citealt{1996ApJ...470L..97H}; \citealt{1999ApJ...522..487H}; \citealt{1999ApJ...519..314H}; \citealt{1997A&A...322L...9L}; \citealt{2004MNRAS.350.1301H}; \citealt{2007ApJ...658L..51C}; \citealt{2008ApJ...679.1390H}; \citealt{2009MNRAS.395.2103M}; \citealt{2010ApJ...710.1310M}; \citealt{2011ApJ...735L..31C}; \citealt{2012ApJ...758..123W}; \citealt{2012ApJ...744...69H}; \citealt{2017MNRAS.469.4763M}; \citealt{2018ApJ...861..127M}; \citealt{2019ApJ...885...99A}), as well as by improving binary population synthesis (BPS; e.g., \citealt{2010A&A...515A..89M}; \citealt{2014MNRAS.440L.101R}; \citealt{2014MNRAS.445.3239P}; \citealt{2014A&A...562A..14T}).
While these efforts have led to significant progress, many uncertainties in stellar physics persist, which influences the outcomes of such studies.
    
Convective overshooting is one of the most significant sources of uncertainty in stellar physics. 
It can significantly affect the internal structure of a star and subsequently its evolution within a binary system. 
The extent of convective overshooting is often parameterized as $\delta_\mathrm{ov} H_\mathrm{p}$, where $\delta_\mathrm{ov}$ is a free parameter and $H_\mathrm{p}$ represents the local pressure scale height. 
From the first principles, it is not possible to determine the distance of the convective overshooting region. 
Instead, $\delta_\mathrm{ov}$ is only constrained through observations (e.g., \citealt{1974ApJ...193..109P}; \citealt{1981A&A....93..136M}; \citealt{1994ApJ...426..165D}; \citealt{2000MNRAS.318L..55R}; \citealt{2012ApJ...761..153Z}; \citealt{2013ApJ...766..118M}; \citealt{2016A&A...592A..15C}; \citealt{2020ApJ...904...22V} ), such as measurements from asteroseismology. 
However, asteroseismic studies have revealed that stars with similar masses can have significantly different convective overshooting parameters. 
For instance, the value of $\delta_\mathrm{ov}$ is $0.2$ for KIC $9812850$ with $M\simeq 1.48 M_\mathrm{\odot}$ \citep{2016ApJ...829...68Y}, $\sim1.4$ for KIC $2837475$ with $M\simeq 1.50 M_\mathrm{\odot}$ \citep{2015MNRAS.453.2094Y}, $0.9-1.5$ for Procyon with $M\simeq 1.50 M_\mathrm{\odot}$ (\citealt{2014ApJ...787..164G}; \citealt{2015ApJ...813..106B}), $\sim 0.6$ for HD $49933$ with $M\simeq 1.28M_\mathrm{\odot}$ \citep{2014ApJ...780..152L}, and $1.7-1.8$ for KIC $11081729$ with $M\simeq 1.27 M_\mathrm{\odot}$ \citep{2015arXiv150800955Y}.
These findings emphasize the significant uncertainties faced when determining convective overshooting.

Convective regions are defined by the Schwarzschild criterion, i.e., they are regions where the fluid elements experience acceleration. 
However, at the convective boundary, although the acceleration of fluid elements is zero, their velocity remains nonzero. 
As a result, fluid elements overshoot the convective boundary, forming a region known as the convective overshooting zone. 
By enhancing the internal mixing processes, convective overshooting allows for the transport of hydrogen-rich material into the hydrogen-burning core, which in turn allows hydrogen to burn more efficiently and increases the mass of the core. 
Consequently, stars with convective overshooting tend to have larger radii at the same evolutionary stage compared to those without convective overshooting. 

In this work we investigate the impact of convective overshooting on the progenitors of SNe Ia in WD + MS systems. 
We describe the methods for detailed binary evolution calculation and present the parameter space that leads to SNe Ia for three different convective overshooting parameters in Sect. \ref{sec:2}. We calculate the SN Ia birth rate for three different convective overshooting parameters in Sect. \ref{sec:3} . 
A discussion of our findings is provided in Sect. \ref{sec:4}, followed by conclusions in Sect. \ref{sec:5}.

\begin{figure}
   \centering
   \includegraphics[width=\hsize]{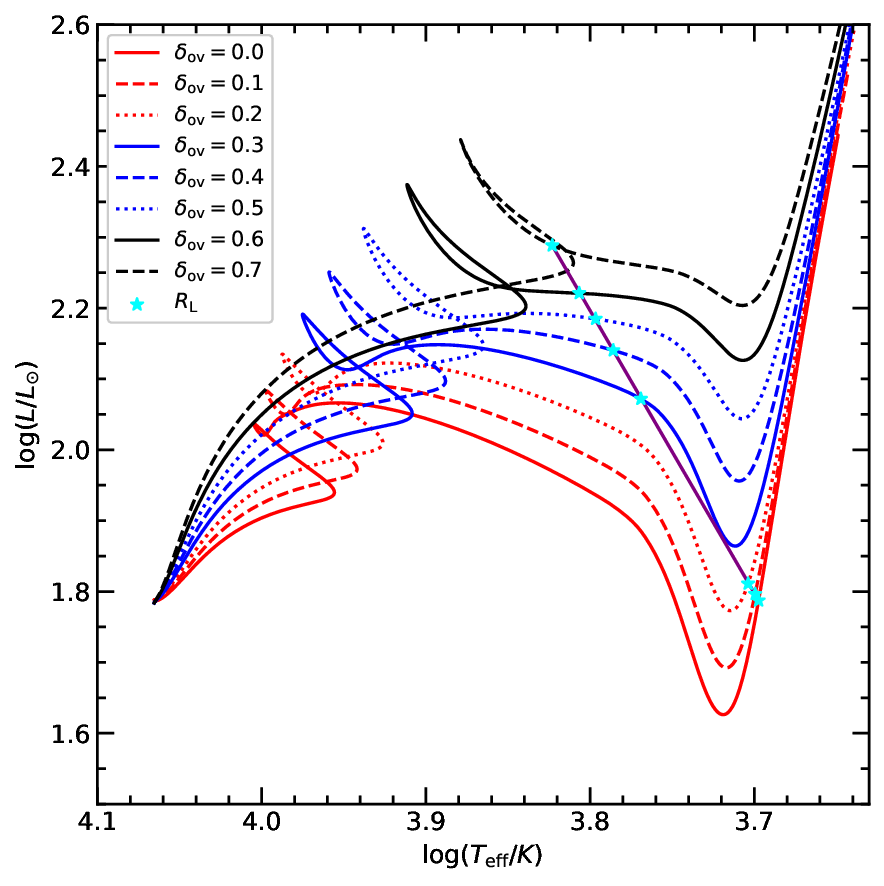}
      \caption{
      Evolution of a star with 2.8$M_\mathrm{\odot}$ in the Hertzsprung--Russell diagram with different $\delta_\mathrm{ov}$.
      The $\delta_\mathrm{ov}$ is set to 0.00, 0.10, 0.20, 0.30, 0.40, 0.50, 0.60, and 0.70. A star symbol indicates the point at which the star radius equals the Roche radius, $R_\mathrm{L}$, where a WD mass of $M_\mathrm{WD}=1.0M_\mathrm\odot$ and orbital periods of $\mathrm{log}(P^\mathrm{i}/\mathrm{days})=0.8$ are assumed. 
      The purple line connecting the star symbols represents the equal-radius line in the Hertzsprung--Russell diagram.
      }
      \label{fig1}
\end{figure}

\section{Binary calculation and results}\label{sec:2}
\subsection{Methods and physics input}\label{sec:2.1}

In this work we calculated the evolution of binary systems using the most updated and flexible stellar evolution code Modules for Experiments in Stellar Astrophysics  (MESA; version r24.08.1; \citealt{2011ApJS..192....3P}; \citealt{2013ApJS..208....4P}; \citealt{2015ApJS..220...15P}; \citealt{2015ApJS..220...15P}; \citealt{2018ApJS..234...34P}; \citealt{2019ApJS..243...10P}; \citealt{2023ApJS..265...15J}).

In our calculations, we adopted a typical Population I composition with a hydrogen mass fraction X = 0.70, helium mass fraction Y = 0.28, and metallicity Z = 0.02 for initial models. 
The convective regions are defined by the Schwarzschild criterion, with the mixing length parameter set to $\alpha = l/H_\mathrm{p}$, representing the ratio of the typical mixing length to the local pressure scale height, at a value of 2 (\citealt{1997MNRAS.289..869P}; \citealt{1997MNRAS.285..696S}). 
We applied the ``step'' scheme for convective overshooting. 
We assumed that the mixing efficiency in the overshooting zones is equal to that at the boundary of the convective zones. 
To explore the impact of convective overshooting on SN Ia progenitors, we set the convective overshooting parameter, $\delta_\mathrm{ov}$, to 0.00, 0.25, and 0.50. 
We focused on WD + MS systems.
Roche lobe overflow (RLOF) is treated using the ``Kolb'' scheme \citep{1990A&A...236..385K}. 
In this scheme, stars have an extended atmosphere, and the mass transfer process is classified as optically thin or optically thick, depending on whether the Roche lobe radius is larger or smaller than the stellar radius.
We treated WD as a point mass without calculating its internal structure. 
When the mass of WD reaches 1.378 $M_\mathrm{\odot}$, it is assumed that a SN Ia explosion occurs.

We employed the common-envelope wind (CEW) model (\citealt{2017MNRAS.469.4763M}), which is a new version of the SD model, as our progenitor model. 
The CEW model presents several advantages, notably its ability to operate effectively across any metallicity.
Here, we provide a brief introduction to the CEW model. 
For detailed information, please refer to \citet{2017MNRAS.469.4763M}. 
In the CEW model, when the mass transfer rate, $|\dot{M}_\mathrm{2}|$, exceeds the critical rate, $\dot{M}_\mathrm{cr}$, a common envelope (CE) forms.
The CE is divided into two regions: the inner region, which corotates with the binary system, and the outer region, which exhibits differential rotation.
The differential rotation in the outer regions continuously extracts orbital angular momentum from the inner binary. Depending on the presence or absence of a CE, the accretion rate of the WD, $\dot{M}_\mathrm{WD}$, can be expressed as follows:
\begin{equation}
   \left.\dot{M}_{\mathrm{WD}}=\left\{\begin{array}{ll}\dot{M}_\mathrm{cr},&\mathrm{if \ CE \ exist,}
      \\ \eta_\mathrm{He}\eta_\mathrm{H}|\dot{M}_2|,&\mathrm{if \ CE \ does \ not \ exist,}\end{array}\right.\right.
      \label{eq:1}
   \end{equation}
where $\eta_\mathrm{H}$ and $\eta_\mathrm{He}$ are the mass accumulation efficiencies for hydrogen burning and helium flashes, respectively, and $|\dot{M}_2|$ is the mass transfer rate.
The  critical accretion rate, $\dot{M}_\mathrm{cr}$, is expressed as \citep{1999ApJ...519..314H}\begin{equation}
   \dot{M}_{{\mathrm{cr}}}=5.3\times10^{-7}\frac{(1.7-X)}{X}(M_{{\mathrm{WD}}}-0.4) M_{\odot} \mathrm{yr}^{-1},
   \label{eq:2}
\end{equation}
where X is the hydrogen mass fraction and $M_\mathrm{WD}$ is the mass of the accreting WD (in $M_\mathrm{\odot}$).

\subsection{Single star evolution tracks}\label{sec:2.2}

We conducted a test to investigate the effect of the different convective overshooting parameters ($\delta_\mathrm{ov}$ = 0.00, 0.10, 0.20, 0.30, 0.40, 0.50, 0.60, and 0.70) on the evolution of single star with a mass of 2.8 $M_\mathrm{\odot}$ (Fig. \ref{fig1}).
The evolutionary tracks are obviously affected by the convective overshooting. 
The increase in $\delta_{\mathrm{ov}}$ enhances mixing, consistently transporting hydrogen-rich material from the outer layers to the core, and the lifetime of a star on the MS increases.
The greater the amount of fuel in the core, the more massive the core becomes at the same evolutionary stage, which leads to a higher core temperature. 
To maintain equilibrium, the star expands, which results in an increase in radius and luminosity, while the effective temperature decreases.

For the same initial binary parameters ($M_\mathrm{WD}=1.0M_\mathrm\odot$ and  $\mathrm{log}(P^\mathrm{i}/\mathrm{days})=0.8$), for the above reasons, the star with a large $\delta_\mathrm{ov}$ fills its Roche lobe at an earlier evolutionary stage (see Fig. \ref{fig1}). 
If the star fills its Roche lobe during the red giant branch (RGB) phase, the system will enter dynamical instability. 
This suggests that convective overshooting significantly influences the parameter space that leads to SNe Ia.

\begin{figure}[h!]
   \centering
   \includegraphics[width=\linewidth]{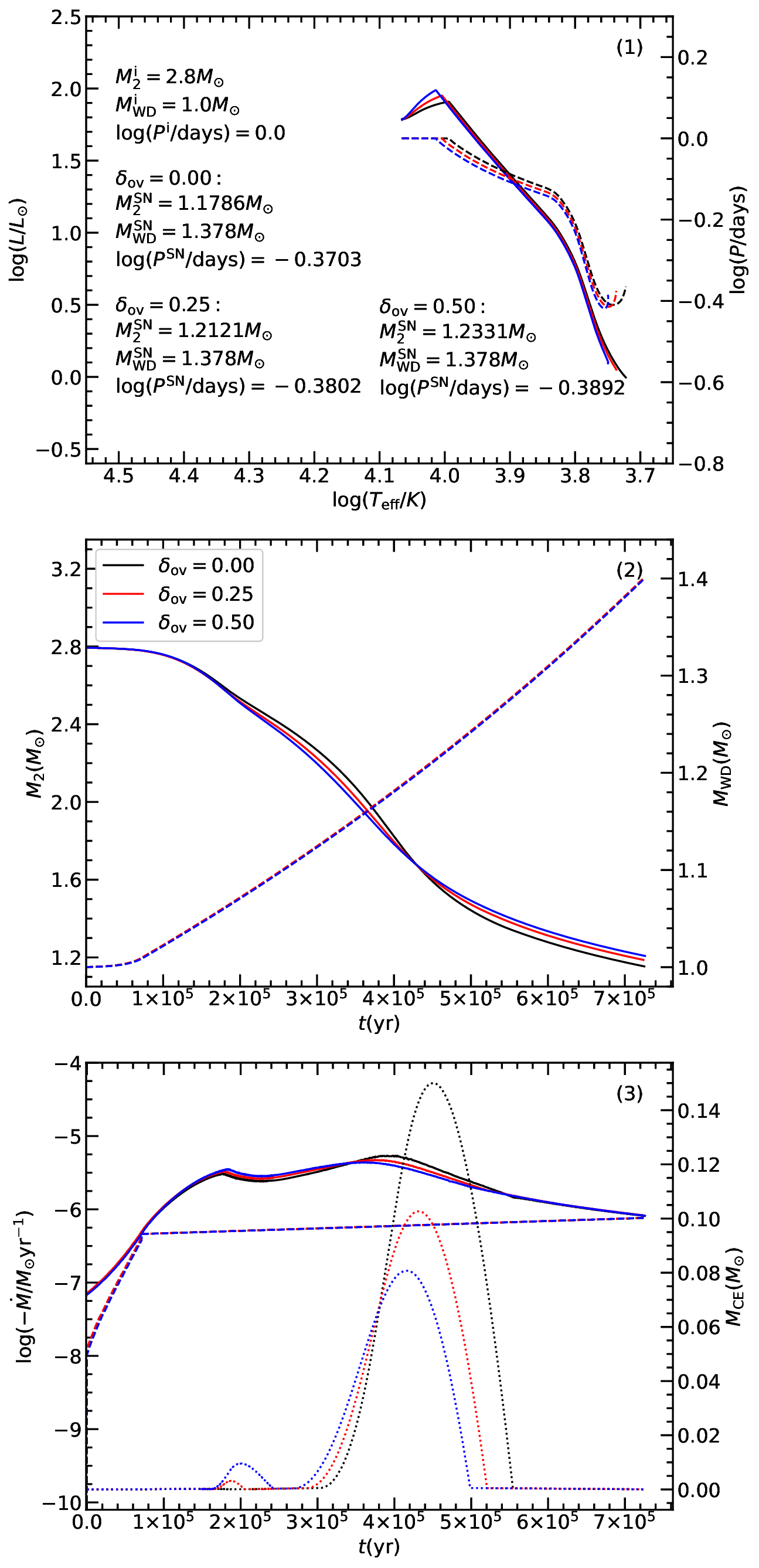}
      \caption{Example of binary evolution. 
      In panel (1), the evolutionary tracks of the donor stars are represented by solid lines and the orbital period evolution with dashed curves. 
      In panel (2), the solid and dashed lines indicate the donor star mass ($M_2$) and the WD mass ($M_\mathrm{WD}$), respectively. 
      In panel (3), the solid, dashed, and dotted lines correspond to the mass transfer rate ($\dot{M}_2$), the mass growth rate of the CO WD ($\dot{M}_\mathrm{WD}$), and the mass of the CE ($M_\mathrm{CE}$), respectively. 
      The black, red, and blue lines represent $\delta_\mathrm{ov} = 0.00$, $0.25$, and $0.50$, respectively. 
      The initial binary parameters and the parameters at the time of the SN Ia explosion are also provided in panel (1).
      }
      \label{fig2}
\end{figure}

\subsection{Binary evolution results}\label{sec:2.3}

\begin{figure}[h!]
   \centering
   \includegraphics[width=\hsize]{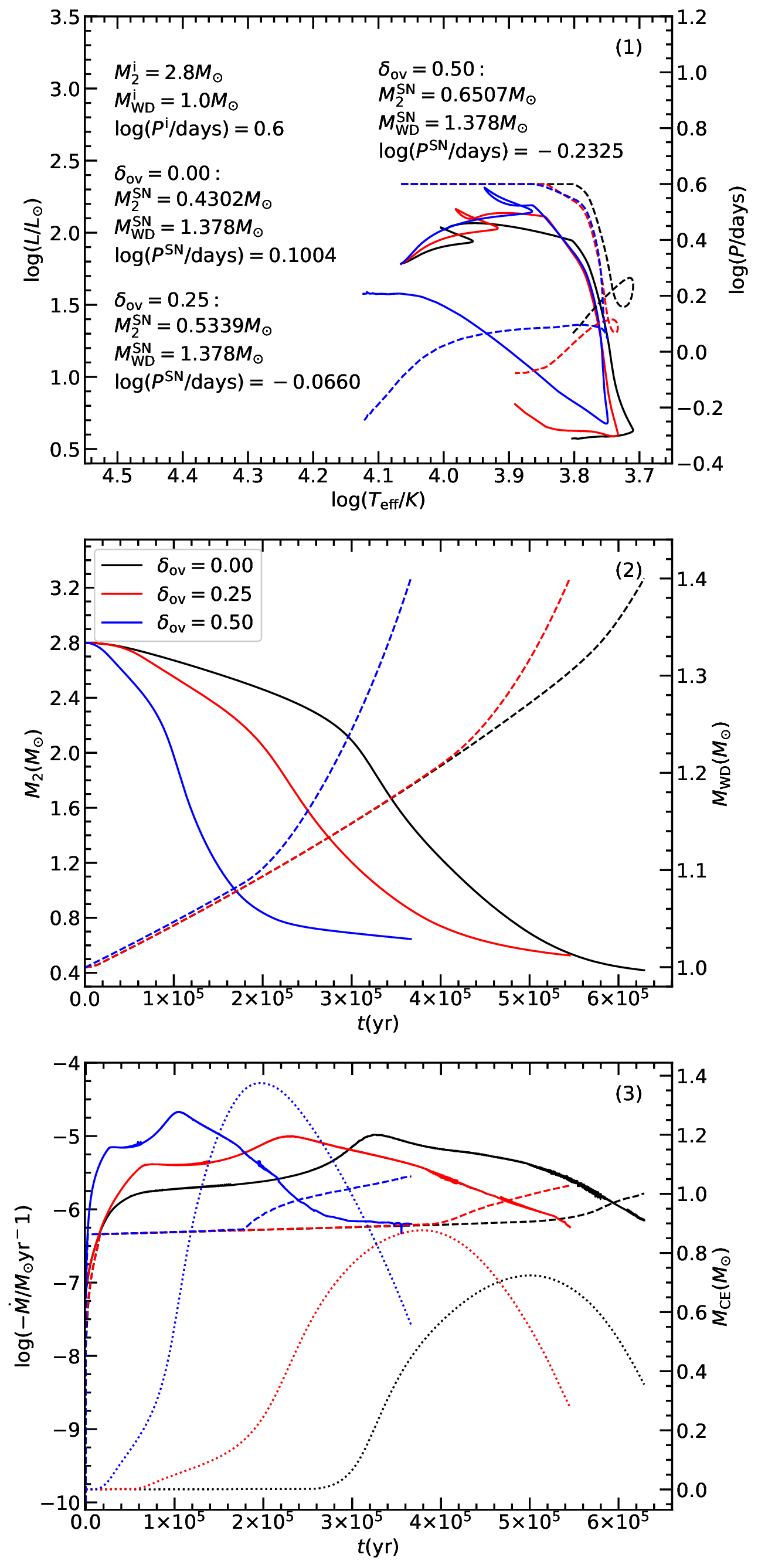}
      \caption{
      Same as Fig. \ref{fig2} except the initial orbital period is different and the donor initiates RLOF in the HG.}
      
      \label{fig3}
\end{figure}

Figures \ref{fig2} and \ref{fig3} present two representative examples from our binary evolution calculations. 
These figures show the evolutionary track of the donor star in the Hertzsprung--Russell diagram, the orbital period evolution, the mass of the WD and donor star, $M_\mathrm{WD}$ and $M_\mathrm{2}$, the mass transfer rate, $|\dot{M}_\mathrm{2}|$, and the accretion rate of the WD, $\dot{M}_\mathrm{WD}$. Figure \ref{fig2} shows the evolution of a binary system with initial parameters: initial donor star mass $M^\mathrm{i}_\mathrm{2}=2.8M_\mathrm{\odot}$, initial  WD mass $M_\mathrm{WD}=1.0M_\mathrm{\odot}$, and initial orbital period $\mathrm{log}(P^\mathrm{i}/\mathrm{days})=0.0$. 
The donor star fills its Roche lobe on the MS, and the system undergoes Case A RLOF. 
The mass transfer rate quickly exceeds the critical accretion rate, which leads to the formation of a common envelope (CE) and the system entering the CE phase. 
The WD gradually increases its mass until it reaches $M^\mathrm{SN}_\mathrm{WD}=1.378M_\mathrm{\odot}$; at this moment it is assumed to explode as a SN Ia. 
At this point, the donor stars mass are $M^\mathrm{SN}_\mathrm{2}=1.1786M_\mathrm{\odot},1.2121M_\mathrm{\odot},1.2331M_\mathrm{\odot}$, and the orbital period is $\mathrm{log}(P^\mathrm{SN}/\mathrm{days})=-0.3703,-0.3802$, and$-0.3892$ for $\delta_\mathrm{ov}=0.00,0.25$, and $0.50$, respectively.
The figure shows that the evolution of the orbital period, the variations in the mass of the WD and the donor star, and the mass transfer rate are similar for different values of $\delta_\mathrm{ov}$; the influence of the different $\delta_\mathrm{ov}$ on the binary evolution is not significant if mass transfer occurs on MS.

Figure \ref{fig3} shows another example of an initial system, where the initial parameters are [$M^\mathrm{i}_\mathrm{2}$, $M_\mathrm{WD}$, $\mathrm{log}(P^\mathrm{i}/\mathrm{days})$]=[$2.8 M_\mathrm{\odot}$, $1.0 M_\mathrm{\odot}$, 0.6].
When the mass of the WD reaches $M^\mathrm{SN}_\mathrm{WD} = 1.378M_\mathrm{\odot}$, the binary parameters are as follows: $M^\mathrm{SN}_2 = 0.4302M_\mathrm{\odot}$, $0.5339M_\mathrm{\odot}$, and $0.6507M_\mathrm{\odot}$, $\mathrm{log}(P^\mathrm{SN}/\mathrm{days}) = 0.1004$, $-0.0660$, and $-0.2325$, for $\delta_\mathrm{ov} = 0.00$, $0.25$, and $0.50$, respectively.
The main difference from the previous example is that the  RLOF occurs in the Hertzsprung gap (HG), and the system undergoes the early Case B RLOF.
From the figure, the evolution of the orbital period, the WD, the donor star, and the mass transfer rate are significantly different for different values of $\delta_\mathrm{ov}$. 
Specifically, the mass transfer rate increases as $\delta_\mathrm{ov}$ increases, which results in earlier WD mass growth (see Fig. \ref{fig3}, middle panel). 

\begin{figure*}
   \centering
   \resizebox{14.54cm}{21.8cm}
   {\includegraphics{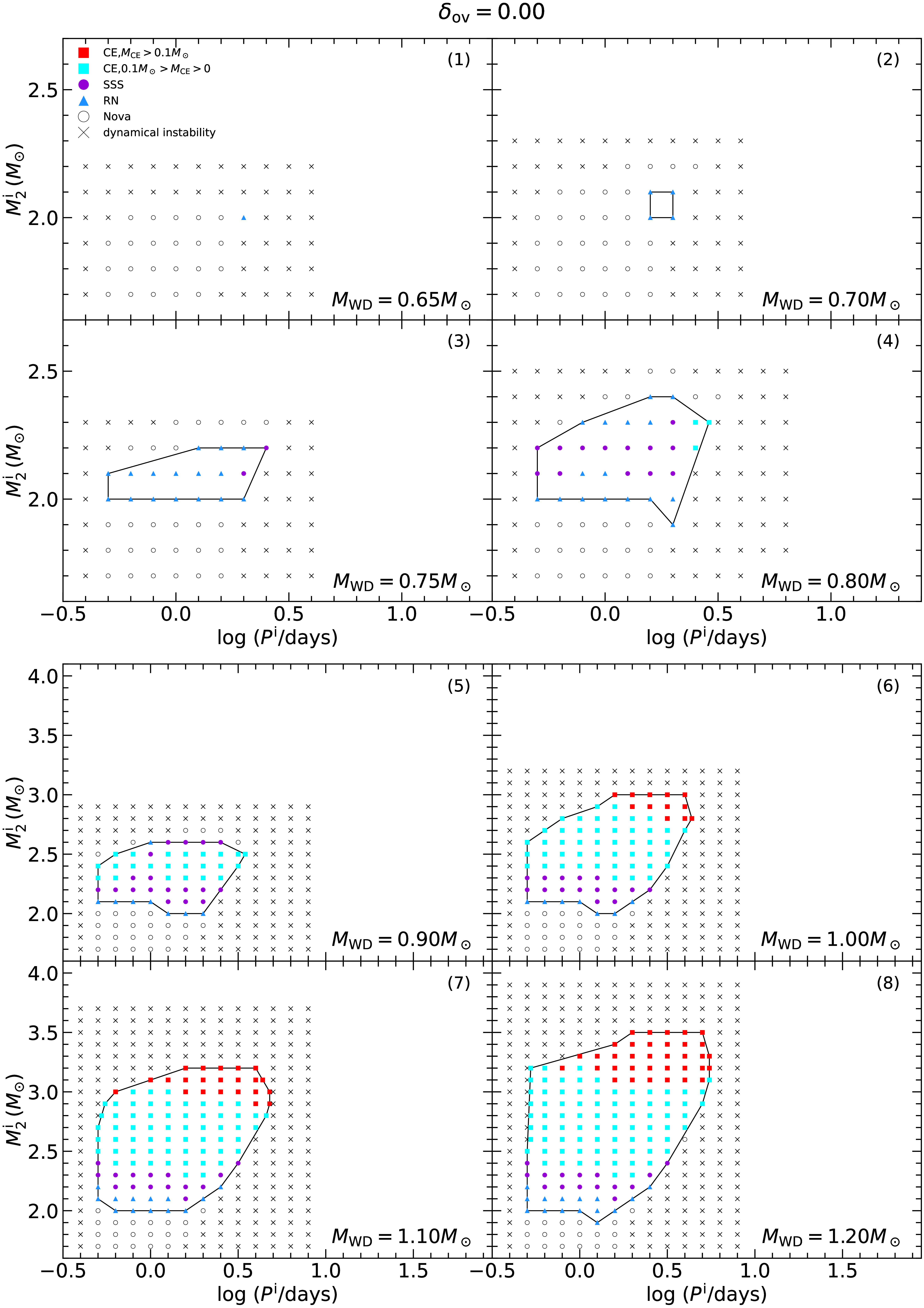}}
      \caption{Final binary evolution calculations for $\delta_\mathrm{ov} = 0.00$, shown in the initial orbital period-secondary mass ($\mathrm{log}P^\mathrm{i}, M_\mathrm{2}^\mathrm{i}$) plane, where $P^\mathrm{i}$ represents the initial orbital period and $M_\mathrm{2}^\mathrm{i}$ is the initial mass of the donor star (for various initial WD masses, as indicated in each panel). 
      Filled squares indicate SN Ia explosions during the CE phase (the red indicates $M_\mathrm{CE} > 0.1 M_\mathrm{\odot}$, the blue $ 0.1 M_\mathrm{\odot} >M_\mathrm{CE} >0 $). 
      Filled circles denote SN Ia explosions in the SSS phase ($\dot{M}_\mathrm{cr} \geq |\dot{M}_2| \geq \frac{1}{2} \dot{M}_\mathrm{cr}$ and $M_\mathrm{CE} = 0$), and filled triangles in the RN phase ($\frac{1}{2} \dot{M}_\mathrm{cr} > |\dot{M}_2| \geq \frac{1}{8} \dot{M}_\mathrm{cr}$ and $M_\mathrm{CE} = 0$). 
      Open circles represent systems that experience nova explosions, preventing the CO WD from reaching 1.378 $M_\odot$ ($|\dot{M}_2| < \frac{1}{8} \dot{M}_\mathrm{cr}$ and $M_\mathrm{CE} = 0$), while crosses show systems 
      that are unstable due to dynamical mass transfer.
      }
      \label{fig4}
\end{figure*}

\begin{figure*}
   \centering
   \resizebox{13cm}{20cm}
   {\includegraphics{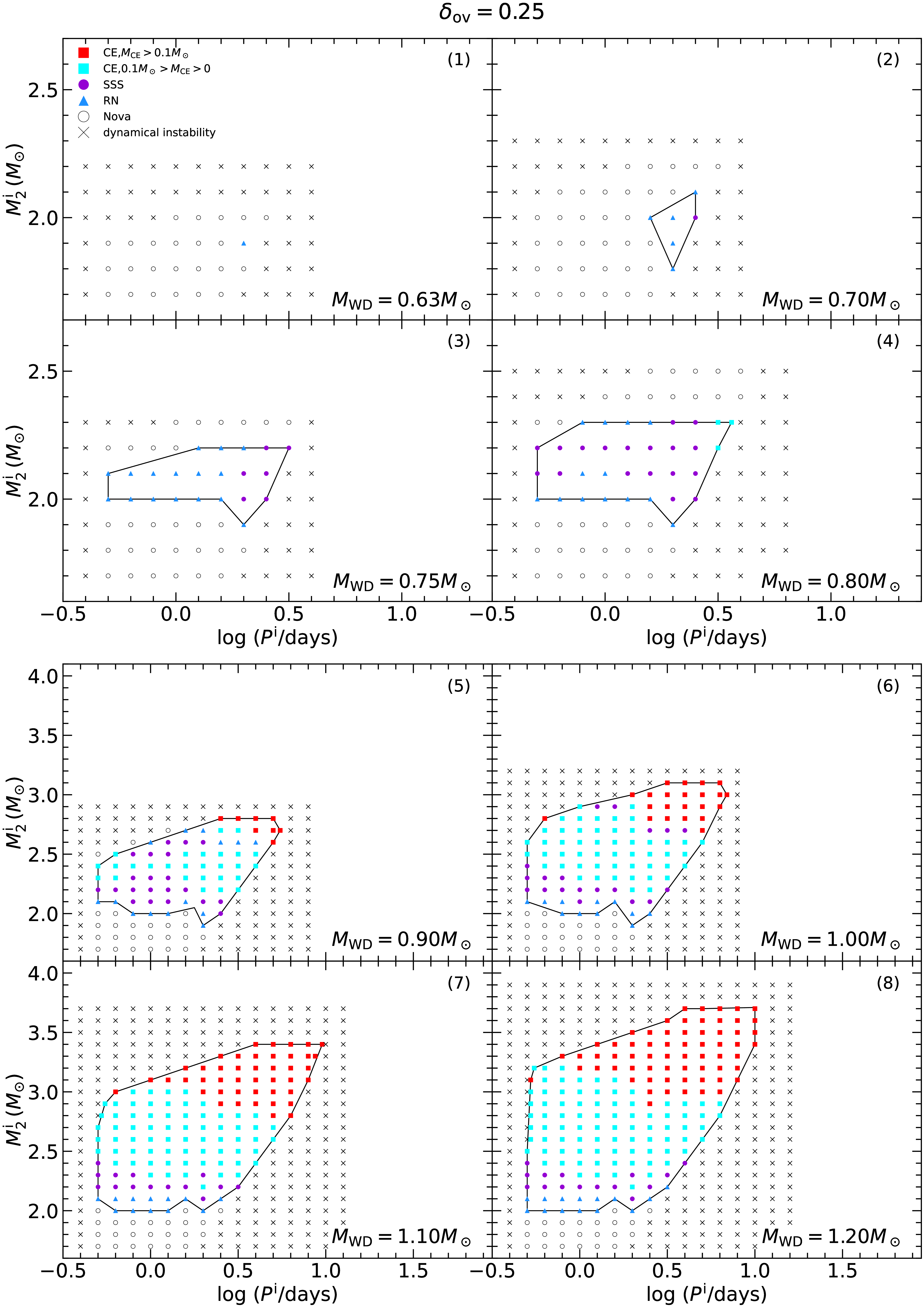}}
      \caption{Same as Fig. \ref{fig4} but for $\delta_\mathrm{ov}=0.25.$}
      \label{fig5}
\end{figure*}

\begin{figure*}
   \centering
   \resizebox{13cm}{20cm}
   {\includegraphics{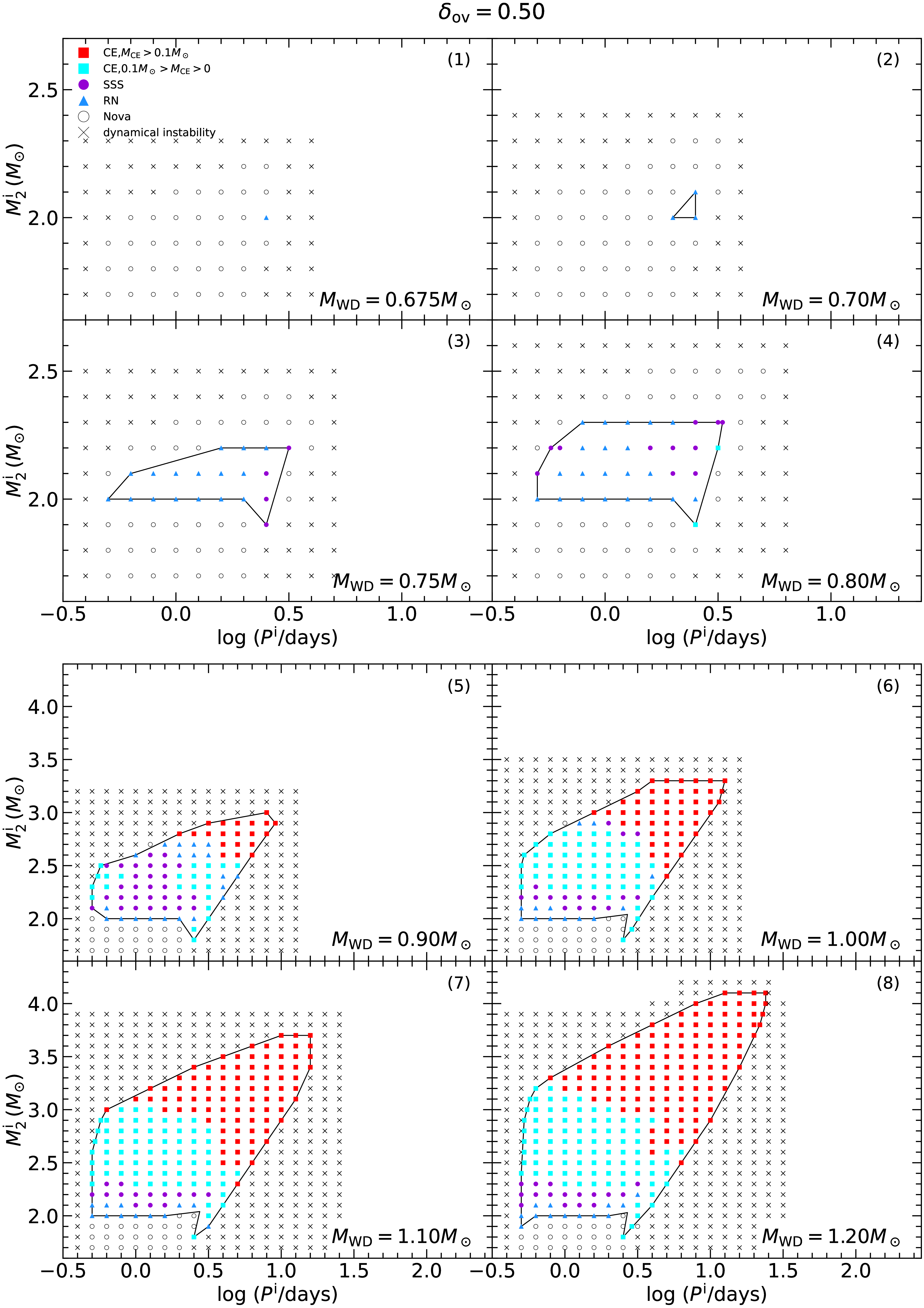}}
      \caption{Same as Fig. \ref{fig4} but for $\delta_\mathrm{ov}=0.50.$}
      \label{fig6}
\end{figure*}

To compare the effects of different $\delta_\mathrm{ov}$ on progenitor systems, we summarize the final outcomes of binary evolution calculation in the initial orbit-initial secondary mass ($\mathrm{log}P^\mathrm{i}, M_\mathrm{2}^\mathrm{i}$) plane (Figs. \ref{fig4}, \ref{fig5} and \ref{fig6}). 
From the three figures for different $\delta_\mathrm{ov}$, CO WDs can reach a mass of 1.378 $M_\odot$ during the CE phase (filled squares), the supersoft X-ray source (SSS) phase (filled circles), or the recurrent nova (RN) phase (filled triangles), as shown in \citet{2017MNRAS.469.4763M}. 
However, due to strong flashes (open circles) or unstable dynamical mass transfer (crosses), many CO WDs fail to reach the mass of 1.378 $M_\odot$. 
Additionally, more systems with large $\delta_\mathrm{ov}$ explode during the CE phase than those with small $\delta_\mathrm{ov}$.

The left boundaries of the contours are decided by the condition at which the RLOF starts when the donor is still unevolved (i.e., on the zero-age MS). 
Beyond the right boundary, the donor evolves onto the RGB, where the system undergoes a dynamically unstable mass transfer. 
The upper boundaries represent regions with very large mass ratios q ($\frac{M_\mathrm{2}}{M_\mathrm{1}}$), which are prone to delayed dynamical instability. 
The lower boundaries arise from the requirement that the mass transfer rate must be sufficiently high for the WD to be able to grow, and the donor mass must be large enough to supply enough material for the WD to reach the $M_\mathrm{Ch}$.

\begin{figure*}
   \centering
   \resizebox{13cm}{10cm}
   {\includegraphics{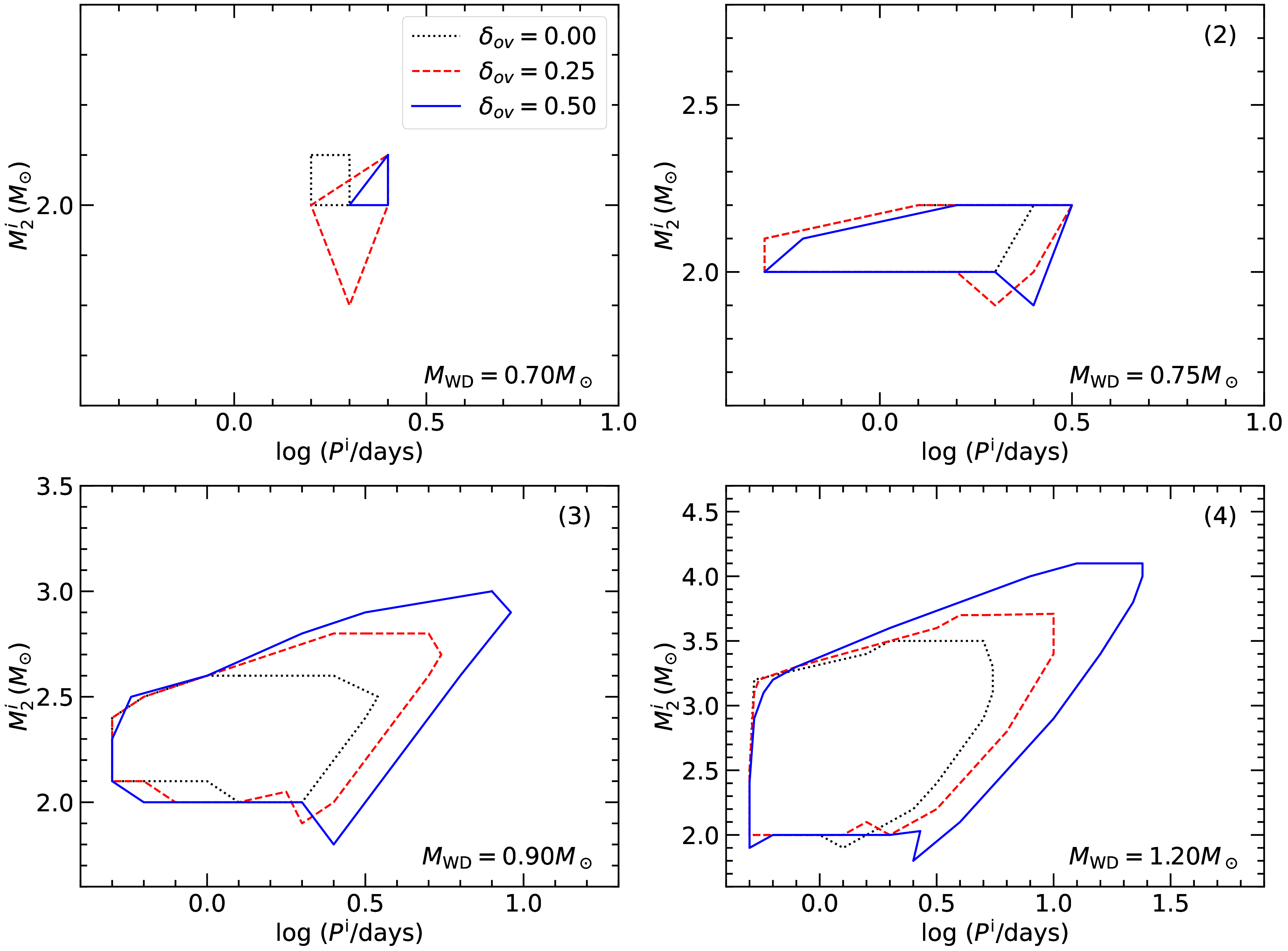}}
      \caption{Contours of initial parameters in the ($\log P^\mathrm{i}, M_2^\mathrm{i}$) plane for different WD masses and varying $\delta_\mathrm{ov}$ values, representing the regions where SNe Ia are expected. The initial masses of the WDs are indicated in the lower-right corner of each panel.
      }
      \label{fig7}
\end{figure*}

We show contours in the ($\mathrm{log}P^\mathrm{i}, M_\mathrm{2}^\mathrm{i}$) plane for four initial WD masses ($0.7M_\mathrm{\odot}, 0.75M_\mathrm{\odot}, 0.90M_\mathrm{\odot},1.20M_\mathrm{\odot}$) with three different $\delta_\mathrm{ov}$ in Fig. \ref{fig7}, which allows us to observe the impact of convective overshooting. 
The figure shows that the contours expand as  $\delta_\mathrm{ov}$ increases when the WD mass exceeds 0.7 $M_\mathrm{\odot}$.
The right boundaries are extended by convective overshooting, which allows systems with relatively long initial orbital periods to contribute to the progenitors of SNe Ia (see also Fig. \ref{fig1}).
The star with a larger $\delta_\mathrm{ov}$ has a larger stellar radius at the same evolutionary stage than those with a smaller $\delta_\mathrm{ov}$. Consequently, companions with large $\delta_\mathrm{ov}$ can initiate RLOF relatively early in their evolution for the same initial orbital period.
For example, a companion with a large $\delta_\mathrm{ov}$ can start RLOF in the HG, whereas a companion with a small $\delta_\mathrm{ov}$ would initiate RLOF in the RGB (see also Fig. \ref{fig1}).

The upper boundaries move upward with the increase in $\delta_\mathrm{ov}$, which suggests that convective overshooting raises the critical mass ratio, $q_{\mathrm{crit}}$. Ratios above this maximum will lead to the delayed dynamical instability.

\begin{figure}[h!]
   \centering
   \includegraphics[width=\hsize]{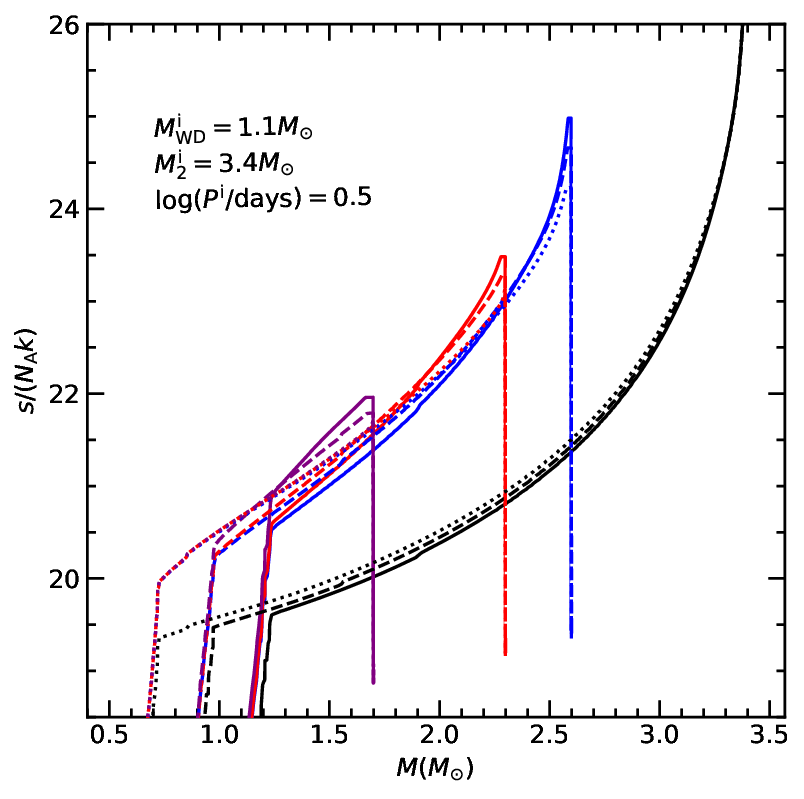}
      \caption{Entropy profiles of the donor star after it loses different amounts of mass for varying $\delta_\mathrm{ov}$. 
      The solid line represents $\delta_\mathrm{ov} = 0.50$, the dashed line $\delta_\mathrm{ov} = 0.25$, and the dotted line $\delta_\mathrm{ov} = 0.00$.
      Black, blue, red, and purple represent the donor losing masses $\bigtriangleup{M} = 0.0M_\mathrm{\odot}$, $\bigtriangleup{M} = 0.8M_\mathrm{\odot}$, $\bigtriangleup{M} = 1.1M_\mathrm{\odot}$, and $\bigtriangleup{M} = 1.7M_\mathrm{\odot}$, respectively.
      The initial parameters are $M^\mathrm{i}_\mathrm{WD}=1.1M_\mathrm{\odot}, M^\mathrm{i}_\mathrm{2}=3.4M_\mathrm{\odot}$, and $\mathrm{log}P^\mathrm{i}\mathrm{(days)}=0.5$.}
      \label{fig8}
\end{figure}

This behavior can be understood from the entropy profile of the donor star with different $\delta_\mathrm{ov}$. 
A donor star with a larger $\delta_\mathrm{ov}$ has a more massive and hotter core, which leads to a radiation envelope with a larger radius but a lower mass, consequently leading to a steeper entropy gradient within the donor star. 
When RLOF begins, a steeper entropy gradient enables the donor star to adjust its radius more efficiently in response to the changing Roche lobe radius. Thus, an increase in $\delta_\mathrm{ov}$ delays the onset of dynamical instability, and in some cases, it may not occur even at the end of the mass transfer phase. Figure \ref{fig8} shows the entropy profiles of the donor star after losing different amounts of mass. 
The figure shows that the entropy gradient increases with the increase in  $\delta_\mathrm{ov}$.
Dynamical instability has already occurred when $\bigtriangleup{M} = 1.1M_\mathrm{\odot}$ for the case of $\delta_\mathrm{ov} = 0.00$.

We have only qualitatively discussed the impact of convective overshooting on delayed dynamical instability. 
However, the exact relationship between the $q_\mathrm{crit}$ and convective overshooting parameters is uncertain. 

Furthermore, as the mass of the WD decreases, the influence of convective overshooting becomes progressively weaker. 
This is because a low WD mass corresponds to a narrower mass range for the companion star, which also has a low mass.
Low-mass stars typically have relatively small convective cores, which reduces the impact of convective overshooting.

However, when the mass of WD is less than 0.75 $M_\odot$, the effect of convective overshooting is not monotonic (as shown in Fig. \ref{fig7}).
The minimum WD mass ($M^\mathrm{min}_\mathrm{WD}$) also follows a non-monotonic trend, with values of $0.65 M_\mathrm{\odot}$, $0.63 M_\mathrm{\odot}$, and $0.675 M_\mathrm{\odot}$ for $\delta_\mathrm{ov} = 0.00$, $\delta_\mathrm{ov} = 0.25$, and $\delta_\mathrm{ov} = 0.50$, respectively, as illustrated in Figs. \ref{fig4}, \ref{fig5}, and \ref{fig6}.
This is because a low-mass WD requires more accreted material to reach $M_\mathrm{Ch}$. 
For a given evolutionary stage of the donor star, the mass transfer rate increases with $\delta_\mathrm{ov}$.
For large $\delta_\mathrm{ov}$, the system has a higher mass transfer rate.
However, WD can only accrete material at a rate up to the $\dot{M}_\mathrm{cr}$, which leads to substantial material loss. 
Although the lost mass can accumulate in the CE, the strong wind from the CE surface causes the envelope to dissipate quickly, which prevents the WD from reaching $M_\mathrm{Ch}$. 
Conversely, for small $\delta_\mathrm{ov}$, the lower mass transfer rate fails to supply sufficient material for the WD. 
By contrast, at $\delta_\mathrm{ov} = 0.25$, the mass transfer rate is moderate -- neither excessively high nor too low -- which enables the WD to accumulate enough mass to reach  $M_\mathrm{Ch}$.

\section{Birth rate estimate}\label{sec:3}

We adopted Eq. (1) from \citet{1984ApJS...54..335I} and followed a similar procedure as \citet{2011ApJ...735L..31C} to enable a direct comparison of the effects of convective overshooting on the SN Ia birth rate in  the Galaxy:
\begin{equation}
   \nu=0.2\mathrm{~yr}^{-1}\Delta q\Delta\log_{10}a\int_{M_\mathrm{A}}^{M_\mathrm{B}}M^{-2.5}dM,
   \label{eq:3}
\end{equation}
where  $M_\mathrm{A}$  and  $M_\mathrm{B}$  represent the typical progenitor mass of WDs at each end of the mass range. 
The integrand is the initial mass function (IMF).
For the quantity  $\Delta\log_{10}a$, we set $\Delta\log_{10}a \approx  \frac{2}{3} \Delta\log_{10}P^\mathrm{i}$ (\citealt{1999ApJ...522..487H}).
The initial range of mass ratios $\Delta q$ for a given WD mass is\begin{equation}
   \Delta q=\frac{M_{\mathrm{u}}}{M_{\mathrm{A}}}-\frac{M_{\mathrm{l}}}{M_{\mathrm{B}}},
\end{equation}
where $ M_\mathrm{u}$  and  $M_\mathrm{l}$ represent the upper and lower limits of the SN Ia progenitor regions, which are directly identified from Figs. \ref{fig4}, \ref{fig5}, and \ref{fig6}.

We calculate the Galactic SN Ia birth rates to be $2.1 \times 10^{-3}\mathrm{yr}^{-1}$, $3.7 \times 10^{-3}\mathrm{yr}^{-1}$, and $3.0 \times 10^{-3} \mathrm{yr}^{-1}$ for $\delta_\mathrm{ov}$ = 0.00, 0.25, and 0.50, respectively (see Tables \ref{table:1}, \ref{table:2}, and \ref{table:3}).
While the parameter space of massive WDs ($\ge 0.75M_\mathrm{\odot}$) increases with $\delta_\mathrm{ov}$, the IMF in Eq. (\ref{eq:3}), $\propto{M^\mathrm{-2.5}}$, indicates that higher-mass systems constitute a smaller fraction of the population. 
Therefore, although the parameter space for massive WDs is larger, they contribute only a small fraction of SNe Ia.
In contrast, SNe Ia are primarily contributed by WDs with a mass around 0.75 $M_\mathrm{\odot}$ (e.g., \citealt{2009MNRAS.395.2103M}).
Meanwhile, the minimum WD mass and the parameter space for low-mass WDs do not vary monotonically with $\delta_\mathrm{ov}$. 
Ultimately, this results in a non-monotonic relation between the computed SN Ia birth rate and $\delta_\mathrm{ov}$.

\begin{table}
   \centering
   \caption{\centering SN Ia birth rate  for $\delta_\mathrm{ov}=0.00.$}   
   \label{table:1}      
   \centering   
   \small    
   \renewcommand{\arraystretch}{1.2}                   
   \begin{tabular}{ >{\centering\arraybackslash}p{1.2cm}
      >{\centering\arraybackslash}p{0.8cm} >{\centering\arraybackslash}p{0.8cm} 
      >{\centering\arraybackslash}p{0.8cm} >{\centering\arraybackslash}p{0.8cm} 
      >{\centering\arraybackslash}p{0.8cm} >{\centering\arraybackslash}p{0.8cm} 
      >{\centering\arraybackslash}p{0.8cm}}        
   \hline\hline   
                 
   $M_\mathrm{WD}/M_\mathrm{\odot}$ & $\bigtriangleup \mathrm{log_\mathrm{10}}a $ & $M_\mathrm{A}/M_\mathrm{\odot}$  & $M_\mathrm{B}/M_\mathrm{\odot}$ & $\bigtriangleup q$ & $\nu / \mathrm{yr^\mathrm{-1}}$ &$\bar{\nu}/\mathrm{yr}^{-1}$ \\   
   
   \hline                        
      0.65-0.7 & 0.07 & 2.42 & 3.20 & 0.2001 & 0.00016& 0.00027\\      
      0.65-0.7 & 0.13 & 2.42 & 3.20 & 0.2414 &0.00039& \\
   \hline
      0.7-0.75 & 0.13 & 3.20 & 3.86 & 0.1396 & 0.00011 & 0.00032\\      
      0.7-0.75 & 0.53 & 3.20 & 3.86 & 0.1709 & 0.00053 & \\
   \hline       
      0.75-0.8 & 0.53 & 3.86 & 4.48 & 0.1226 & 0.00023 & 0.00031\\     
      0.75-0.8 & 0.57 & 3.86 & 4.48 & 0.1967 & 0.00039 & \\
   \hline
      0.8-0.9 & 0.57 & 4.48 & 5.60 & 0.1967 & 0.00045 & 0.00051\\     
      0.8-0.9 & 0.63 & 4.48 & 5.60 & 0.2235 & 0.00056 & \\                   
   \hline
      0.9-1.0 & 0.63 & 5.60 & 6.63 & 0.1625 & 0.00023 & 0.00030\\     
      0.9-1.0 & 0.69 & 5.60 & 6.63 & 0.2340 & 0.00036 & \\ 
   \hline
      1.0-1.1 & 0.69 & 6.63 & 7.58 & 0.1891 & 0.00019 & 0.00021\\     
      1.0-1.1 & 0.72 & 6.63 & 7.58 & 0.2192 & 0.00023 & \\
   \hline
      1.1-1.2 & 0.72 & 7.58 & 8.48 & 0.1863 & 0.00013 & 0.00016\\     
      1.1-1.2 & 0.76 & 7.58 & 8.48 & 0.2377 & 0.00018 & \\
   \hline
   \end{tabular}
   \begin{tablenotes}[flushleft]
      
      \item \textbf{Notes.} For each range of WD masses $M_\mathrm{WD}$, $M_\mathrm{A}$, and $M_\mathrm{B}$ represent the corresponding lower and upper zero-age progenitor masses, respectively. $\Delta q$ and $\Delta \log_{10}a$ denote the typical mass ratio and separation ranges for the regions in Figs. \ref{fig2}, \ref{fig3}, and \ref{fig4} that lead to SNe Ia. The $\nu$ and $\bar{\nu}$ are the calculated SN Ia birth rate, representing the extremes of the distribution applied across the entire range, with $\bar{\nu}$ being a simple mean of these two extremes.
   \end{tablenotes}
\end{table}

\begin{table}
   \centering
   \caption{\centering SN Ia birth rate for $\delta_\mathrm{ov}=0.25.$}   
   \label{table:2}      
   \centering   
   \small    
   \renewcommand{\arraystretch}{1.2}                   
   \begin{tabular}{ >{\centering\arraybackslash}p{1.2cm}
      >{\centering\arraybackslash}p{0.80cm} >{\centering\arraybackslash}p{0.80cm} 
      >{\centering\arraybackslash}p{0.80cm} >{\centering\arraybackslash}p{0.8cm} 
      >{\centering\arraybackslash}p{0.8cm} >{\centering\arraybackslash}p{0.8cm} 
      >{\centering\arraybackslash}p{0.8cm}}     
   \hline\hline   
                 
   $M_\mathrm{WD}/M_\mathrm{\odot}$ & $\bigtriangleup \mathrm{log_\mathrm{10}}a $ & $M_\mathrm{A}/M_\mathrm{\odot}$  & $M_\mathrm{B}/M_\mathrm{\odot}$ & $\bigtriangleup q$ & $\nu / \mathrm{yr^\mathrm{-1}}$ &$\bar{\nu}/\mathrm{yr}^{-1}$ \\   
   \hline                        
      0.63-0.7 & 0.07 & 2.04 & 3.20 & 0.3349 & 0.00050 & 0.00128\\      
      0.63-0.7 & 0.20 & 2.04 & 3.20 & 0.4641 & 0.00207 & \\
   \hline
      0.7-0.75 & 0.20 & 3.20 & 3.86 & 0.1913 & 0.00022 & 0.00050\\      
      0.7-0.75 & 0.60 & 3.20 & 3.86 & 0.2226 & 0.00077 & \\
   \hline       
      0.75-0.8 & 0.60 & 3.86 & 4.48 & 0.1673 & 0.00035 & 0.00036\\     
      0.75-0.8 & 0.64 & 3.86 & 4.48 & 0.1708 & 0.00038 & \\
   \hline
      0.8-0.9 & 0.64 & 4.48 & 5.60 & 0.1744 & 0.00045 & 0.00066\\     
      0.8-0.9 & 0.76 & 4.48 & 5.60 & 0.2860 & 0.00087 & \\                   
   \hline
      0.9-1.0 & 0.76 & 5.60 & 6.63 & 0.2134 & 0.00036 & 0.00043\\     
      0.9-1.0 & 0.83 & 5.60 & 6.63 & 0.2669 & 0.00050 & \\ 
   \hline
      1.0-1.1 & 0.83 & 6.63 & 7.58 & 0.2173 & 0.00026 & 0.00029\\    
      1.0-1.1 & 0.92 & 6.63 & 7.58 & 0.2494 & 0.00033 & \\
   \hline
      1.1-1.2 & 0.92 & 7.58 & 8.48 & 0.2127 & 0.00019 & 0.00021\\  
      
      1.1-1.2 & 0.93 & 7.58 & 8.48 & 0.2522 & 0.00023 & \\
   \hline
   \end{tabular}
   \begin{tablenotes}[flushleft]
      
      \item \textbf{Notes.} Same as Table \ref{table:1} but for $\delta_\mathrm{ov} = 0.25$.
   \end{tablenotes}
\end{table}

\begin{table}
   \centering
   \caption{\centering SN Ia birth rate  for $\delta_\mathrm{ov}=0.50.$}   
   \label{table:3}      
   \centering   
   \small    
   \renewcommand{\arraystretch}{1.2}                   
   \begin{tabular}{ >{\centering\arraybackslash}p{1.3cm}
      >{\centering\arraybackslash}p{0.75cm} >{\centering\arraybackslash}p{0.8cm} 
      >{\centering\arraybackslash}p{0.8cm} >{\centering\arraybackslash}p{0.75cm} 
      >{\centering\arraybackslash}p{0.8cm} >{\centering\arraybackslash}p{0.9cm} 
      >{\centering\arraybackslash}p{0.8cm}}        
   \hline\hline   
                 
   $M_\mathrm{WD}/M_\mathrm{\odot}$ & $\bigtriangleup \mathrm{log_\mathrm{10}}a $ & $M_\mathrm{A}/M_\mathrm{\odot}$  & $M_\mathrm{B}/M_\mathrm{\odot}$ & $\bigtriangleup q$ & $\nu / \mathrm{yr^\mathrm{-1}}$ &$\bar{\nu}/\mathrm{yr}^{-1}$ \\   
   
   \hline                        
      0.675-0.7 & 0.07 & 2.83 & 3.20 & 0.0814 & 0.00003 & 0.00005\\      
      0.675-0.7 & 0.13 & 2.83 & 3.20 & 0.1167 & 0.00007 & \\
   \hline
      0.7-0.75 & 0.13 & 3.20 & 3.86 & 0.1396 & 0.00011 & 0.00040\\      
      0.7-0.75 & 0.60 & 3.20 & 3.86 & 0.1968 & 0.00068 & \\
   \hline       
      0.75-0.8 & 0.60 & 3.86 & 4.48 & 0.1449 & 0.00030 & 0.00033\\     
      0.75-0.8 & 0.61 & 3.86 & 4.48 & 0.1708 & 0.00036 & \\
   \hline
      0.8-0.9 & 0.61 & 4.48 & 5.60 & 0.1744 & 0.00043 & 0.00085\\     
      0.8-0.9 & 0.91 & 4.48 & 5.60 & 0.3486 & 0.00127 & \\                   
   \hline
      0.9-1.0 & 0.91 & 5.60 & 6.63 & 0.2642 & 0.00054 & 0.00063\\     
      0.9-1.0 & 1.00 & 5.60 & 6.63 & 0.3178 & 0.00071 & \\ 
   \hline
      1.0-1.1 & 1.00 & 6.63 & 7.58 & 0.2607 & 0.00037 & 0.00043\\     
      1.0-1.1 & 1.07 & 6.63 & 7.58 & 0.3211 & 0.00050 & \\
   \hline
      1.1-1.2 & 1.07 & 7.58 & 8.48 & 0.2758 & 0.00030 & 0.00034\\     
      1.1-1.2 & 1.19 & 7.58 & 8.48 & 0.3286 & 0.00039 & \\
   \hline
   \end{tabular}
   \begin{tablenotes}[flushleft]
      
      \item \textbf{Notes.} Same as Table \ref{table:1} but for $\delta_\mathrm{ov} = 0.50$.
   \end{tablenotes}
\end{table}

\section{Discussion}\label{sec:4}

\subsection{Uncertainties in our model}
For our model we adopted the ``step'' scheme and assumed that the mixing efficiency in the overshooting zones is equal to that at the boundary of the convective zones. 
However, the treatment of convective overshooting remains highly uncertain. 
Various approaches have been proposed, such as exponential diffusive overshooting (\citealt{2000A&A...360..952H}) 
and turbulent entrainment (\citealt{2007ApJ...667..448M}). These different methods can yield varying results in certain aspects.

Additionally, we simplified our model by assuming that all stars have the same convective overshooting parameter. 
However, previous studies suggest that there could be a certain correlation between stellar mass and convective overshooting, which the $\delta_\mathrm{ov}$ increases with star mass (\citealt{1997MNRAS.285..696S}; \citealt{2016A&A...592A..15C};\citealt{2021A&A...646A.133H}; \citealt{2022ApJ...929..182J}; \citealt{2022A&A...667A..97A}). 
If $\delta_\mathrm{ov}$ increases with stellar mass, it would extend both the right and upper boundaries of the parameter space for massive WD. 
It may also reduce the minimum WD mass required for SNe Ia, which could potentially increase the SN Ia birth rate of the SD model.
However, the relationship remains poorly quantified.

\subsection{Mass transfer rate}\label{sec:4.2}

As previously mentioned, systems with a large $\delta_\mathrm{ov}$ have high mass transfer rates when the donor star is at a given evolutionary stage.
For simplicity, we used the thermal timescale mass transfer rate to reflect the actual mass transfer rate qualitatively:
\begin{equation}
   \dot{M}_{\mathrm{th}}\simeq\frac{\left(M_2^\mathrm{i}-M_1^\mathrm{i}\right)}{\tau_\mathrm{KH}},
\end{equation}
where $M_1^\mathrm{i}$ and $M_2^\mathrm{i}$ are the initial masses of the accretor and donor, respectively, 
and $\tau_\mathrm{KH}$ is the Kelvin-Helmholtz timescale, given by
\begin{equation}
   \tau_{\mathrm{KH}}\simeq\frac{GM_2^2}{2R_2L_2},
\end{equation}
where $G$ is the gravitational constant, and $M_2$, $R_2$, and $L_2$ represent the mass, radius, and luminosity of the donor star, respectively.

\begin{figure}[h!]
   \centering
   \includegraphics[width=\hsize]{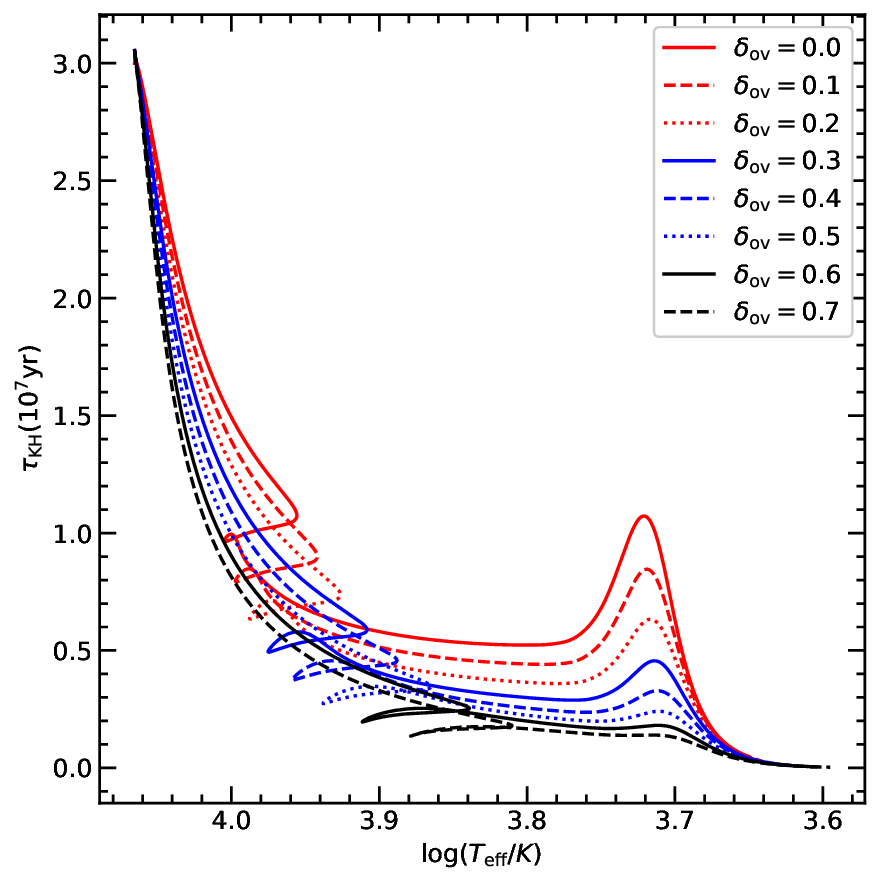}
      \caption{Evolution of the Kelvin-Helmholtz timescale for stars with $2.8M_\mathrm{\odot}$ and different $\delta_\mathrm{ov}$. The $\delta_\mathrm{ov}$ is set to 0.00, 0.10, 0.20, 0.30, 0.40, 0.50, 0.60, and 0.70.
      }
      \label{fig9}
\end{figure}

Figure \ref{fig9} shows the evolution of the Kelvin-Helmholtz timescale for stars with different convective overshooting parameters ($\delta_\mathrm{ov} = 0.00, 0.10, 0.20, 0.30, 0.40, 0.50, 0.60, 0.70$). 
The figure shows that during the MS and HG phases, the $\tau_\mathrm{KH}$ decreases as the star evolves. 
Moreover, the $\tau_\mathrm{KH}$ reduces with an increase in $\delta_\mathrm{ov}$. 
As a result, $\dot{M}_\mathrm{th}$ increases with $\delta_\mathrm{ov}$, which leads systems with large $\delta_\mathrm{ov}$ to exhibit a high mass transfer rate.
It should be noted that for systems with the same initial binary parameters, a large $\delta_\mathrm{ov}$ does not necessarily result in a high mass transfer rate. 
This depends on the evolutionary stage of the donor star when it begins RLOF. 
For instance, in a system with $\delta_\mathrm{ov} = 0.00$, the donor star may start RLOF in the HG, whereas in a system with $\delta_\mathrm{ov} = 0.50$, the donor star begins RLOF during the MS. 
At this point, the thermal timescale for $\delta_\mathrm{ov} = 0.00$ could be shorter than that for $\delta_\mathrm{ov} = 0.50$, which results in a higher mass transfer rate for the $\delta_\mathrm{ov} = 0.00$ system.

Additionally, excessive mass transfer does not necessarily increase the birth rate of SNe Ia. 
This is because the WD can only accrete material up to the critical accretion rate $\dot{M}_\mathrm{cr}$, any additional material is lost from the system. Conversely, if the mass transfer rate is too low, the donor cannot provide sufficient material for the WD to reach the Chandrasekhar mass ($M_\mathrm{ch}$).
Only an optimal mass transfer rate allows the birth rate of SNe Ia to reach its maximum. 
Therefore, to address the issue of the birth rate in the SD, it is crucial to investigate the accretion process of WDs. Currently, the accretion rate of WDs remains a significant uncertainty (e.g., \citealt{2012NewAR..56..122W}; \citealt{2015A&A...584A..37W}; \citealt{2018RAA....18...49W}).

\subsection{SNe Ia-CSM}\label{sec:4.3}
Type Ia supernovae that interact with the  circumstellar medium (CSM; hereafter   SNe Ia-CSM) are characterized by narrow H/He emission lines, and they maintain a high luminosity over a long period ( > 1yr). 
The presence of narrow H/He lines and high luminosity suggests that the progenitor systems underwent significant mass loss, producing a massive, H/He-rich CSM before the supernova explosion (e.g., \citealt{2013MNRAS.436..222M}; \citealt{2019MNRAS.487.2372V}; \citealt{2023ApJ...944..204U}).
SNe Ia-CSM are most frequently found in late-type and irregular galaxies, which indicates that they likely arise from relatively young stellar populations (\citealt{2013ApJS..207....3S}).

In the CEW model, if a WD explodes during the CE phase, and the mass of CE exceeds 0.1 $M_\mathrm{\odot}$, hydrogen lines can be directly detected in the spectra \citep{2018ApJ...861..127M}. 
A massive CE is typically formed in systems with massive companions. 
These massive stars have short evolutionary timescales, which leads to shorter delay times. 
This aligns with the observation that SNe Ia-CSM are associated with relatively young stellar populations.

Figures \ref{fig4}, \ref{fig5}, and \ref{fig6} show that a large $\delta_\mathrm{ov}$ significantly increases the number of systems capable of forming a CE with  $M_\mathrm{CE} > 0.1 M_\mathrm{\odot}$.
This markedly enhances the predicted proportion of SNe Ia-CSM.
We calculated the birth rates and proportions of SNe Ia-CSM for different $\delta_\mathrm{ov}$, $\delta_\mathrm{ov}=0.00, 0.25$, and  $0.50 $:  
$1.1 \times 10^{-4}, 4.20\% $;  $3.1 \times 10^{-4}, 8.40\% $; and $6.9 \times 10^{-4}, 22.89\%$, respectively. 
Therefore, convective overshooting significantly increases the theoretically predicted proportion of SNe Ia-CSM.

However, most SNe Ia lack H/He lines. \citet{2012Sci...337..942D} estimated that the fraction of SNe Ia-CSM that exhibit prominent circumstellar interaction near maximum light ranges from 0.1\% to 1\%. 
Although this fraction could be higher if cases with weaker H signatures have been overlooked or if some cases with stronger signatures have been misclassified as Type IIn SNe, SNe Ia-CSM remain rare. 
However, theoretical calculations predict a significantly higher proportion of SNe Ia-CSM. For instance, the proportion of SNe Ia-CSM is as high as 22.89\% for $\delta_\mathrm{ov}$= 0.50.
This might require a time delay between the WD reaching  $1.378~M_\mathrm{\odot}$ and the subsequent supernova explosion. 
The delay would need to be long enough for the CE to dissipate.
One possible mechanism is the spin-up/spin-down model (e.g., \citealt{2011ApJ...730L..34J}; \citealt{2011ApJ...738L...1D}; \citealt{2012ApJ...759...56D}; \citealt{2012ApJ...744...69H};  \citealt{2014MNRAS.445.2340W}). 
During this spin-down phase, the CE can dissipate. 
\citet{2017MNRAS.469.4763M} discuss this problem in detail, suggesting that no more than 1.7 in 100 SNe Ia belong to the SN Ia-CSM class if the spin-down timescale exceeds $10^5$ years.
However, the timescale for the WD's spin-down remains uncertain (e.g., \citealt{2011ApJ...730L..34J}; \citealt{2011ApJ...738L...1D}; \citealt{2012ApJ...759...56D}; \citealt{2012ApJ...744...69H}; \citealt{2013ApJ...778L..35M}; \citealt{2014MNRAS.445.2340W}).

However, the birth rate and the proportion of SNe Ia-CSM could be overestimated by Eq. (\ref{eq:3}) because the calculated region is a rectangular area that includes some points where SNe Ia cannot be produced. 
However, this does not affect the trends in either the SN Ia birth rate or the proportion of SNe Ia-CSM for different $\delta_\mathrm{ov}$.

\section{Conclusions}\label{sec:5}

We have studied the impact of convective overshooting on the SD model of SNe Ia. 
We obtained the parameter space ($\log P^\mathrm{i}, M^\mathrm{i}$) for SNe Ia by calculating binary evolution in detail.
Subsequently, we calculated the SN Ia birth rate for different convective overshooting parameters. 
The main conclusions of our study are as follows:
\begin{enumerate}[label=\arabic*)] 
   \item Convective overshooting expands the parameter space for systems with massive WDs ($\ge 0.75 M_\mathrm{\odot}$). 
   However, the minimum WD mass and the parameter space for low-mass WDs do not vary monotonically with $\delta_\mathrm{ov}$, which results in a non-monotonic trend when calculating the SN Ia birth rate.  
   \item The upper boundaries of the initial parameter space are extended by convective overshooting, which enables systems with massive companions to contribute to the production of SNe Ia.
   Consequently, the $q_\mathrm{crit}$ increases with $\delta_\mathrm{ov}$. 
   \item The mass transfer rate in the system increases with $\delta_\mathrm{ov}$ when the donor is in the same evolutionary stage.

   \item The $\delta_\mathrm{ov}$ has a notable impact on the initial parameter space for SNe Ia-CSM, which increases with $\delta_\mathrm{ov}$.

\end{enumerate}

\begin{acknowledgements}
   We are grateful to Zhengwei Liu, Zhenwei Li, Shunyi Lan, Boyang Guo, Lifu Zhang, Jinxiao Luo, Junda Zhou, and Tan Liu for their fruitful discussions.
   This work is supported by the Strategic Priority Research Program of the Chinese Academy of Sciences (grant Nos. XDB1160303, XDB1160000), the National Natural Science Foundation of China (Nos. 12288102 and 12333008), and the National Key R\&D Program of China (No. 2021YFA1600403). X.M. acknowledges support from Yunnan Fundamental Research Projects (Nos. 202401BC070007 and 202201BC070003), International Centre of Supernovae, Yunnan Key Laboratory (No. 202302AN360001), the Yunnan Revitalization Talent Support Program Science \& Technology Champion Project (No. 202305AB350003), and the science research grants from the China Manned Space Program with grant no. CMS-CSST-2025-A13.
\end{acknowledgements}

\bibliographystyle{aa} 
\bibliography{aa} 

\end{document}